# Rigid Material on Top of a Compliant Flooring Effectively Reduces the Impact Force In The Event of a Forward Fall

Nader Rajaei, *Member, IEEE*, Saeed Abdolshah, *Member, IEEE*, Yasuhiro Akiyama, *Member, IEEE*, Yoji Yamada *Member, IEEE*, Shogo Okamoto *Member, IEEE*,

*Abstract*— The biomechanical studies have proposed several forward fall arresting strategies to reduce the impact forces in a human-robot collaboration environment. A proposed strategy is using a compliant flooring for an environment because it can reduce the stiffness of the ground. In this study, we proposed that if a rigid layer is mounted on a compliant flooring (double-layer flooring), the impact force is further reduced. In order to investigate this goal, we designed two subjective laboratory experiments. In experiment 1, the subjects were instructed falling to the ground where was covered by a single layer of compliant material (a foam pad). In experiment 2, the subjects fell on double-layer flooring when one rigid layer (a wood surface) was mounted on the compliant layer. The impact forces were measured for two short forward fall heights onto the outstretched hand. The results showed that the profile of the impact forces consists of two peaks. The first peak has a higher magnitude, and it is followed by the second peak with a lower magnitude. Comparing the magnitude of the first peak between two experiments shows a reduction of the impact force in experiment 2. In contrast, the magnitudes of second peaks are identical for both experiments. Therefore, we concluded that using a double-layer flooring, i.e.one rigid layer and one compliant layer can be an effective strategy for considerably reducing the impact force, and preventing the wrist fractures during the forward fall.

## I. INTRODUCTION

Fall is a common incident that can seriously threaten the health of humans. A fall leads to physical consequences, or psychophysical, and social consequences, or both [1]-[4]. The physical consequences of fall cause either fatal or non-fatal injuries and exert a significant financial burden on the healthcare system. For example, it has been estimated that in the United States direct cost of occupational fall-related injuries reaches $43.8 billion by the year 2020 [5]. Therefore, it is important to reduce the number of fall-related injuries by effective methods, especially in the occupational environments [6] where humans and robots collaborate and a collision between them may lead to a fall.

There are several types of fall in which a forward fall is one of the most frequent one [7]. It occurs mainly due to an unexpected perturbation of human balance or a failure in his/her recovery strategy[8] when tripping or slipping occurs during their walking [9]-[10]. The forward fall leads to most cases a severe injury or a bone fracture in the upper extremity. The Colles' fracture (distal radius fracture) is one of the common forward fall-related fractures [11]-[13] because humans spontaneously use their hands to avoid injury to the head or trunk when a forward fall occurs. A worst-case forward fall scenario is when the human descends with the outstretched hand position which leads to approximately 90% of the Colles' fracture [14].

A wide range of biomechanical studies has attempted to identify the range of forces that lead to a distal radius fracture using cadaveric hands (the *in vitro* studies) [14]-[17]. Moreover, the *in vivo* studies, the profile of impact forces applied to the hand was identified during low-height forwards fall onto the outstretched hand [18]-[19]. By using mathematical models, the real impact forces during a forward fall from a standing position were estimated [18], [20]-[21] where the *in vivo* studies have shown limitations due to safety considerations for the subjects when falling from a large height.

In addition, the attention of the biomechanical researchers has also attracted to find the effective biomechanical factors that can probably reduce the impact force in the event of a forward fall. A recent study showed that the natural reaction of the human during a forward fall has an important role in preventing injuries [22]. Other studies showed that using a typical fall arresting strategy is an effective method for reducing the fall-related fractures [23]-[26].

One of the newfound methods for preventing fall-related injured is using a compliant flooring [27]. A compliant flooring is a floor covering to reduce the stiffness of the ground. In one early study, the impact forces applied to hands was measured when the subjects fell on a rigid surface and fell on several compliant surfaces with different stiffness [28]. The result showed that the compliant surfaces led to a considerable reduction in the impact force in comparison with the impact force on a rigid surface. Another additional study investigated the magnitude of impact forces when the young adults fall for different playground surface materials. The study results showed a moderate but considerably reduction in the peak of forces for lower stiff materials in comparison with rigid materials [29].

In this study, we investigated the effect of using a double-layer flooring on the magnitude of the impact forces during a forward fall. In order to examine this, a series of subjective laboratory experiments were conducted. We measured the impact forces during fall onto the outstretched hand position from short fall heights to avoid the risk of a fracture of the subjects' hand. We compared the impact forces when the subjects fall on only one layer of a compliant surface and when they fall on a double-layer surface in which a rigid layer is mounted on the compliant layer. We hypothesize that the

N. Rajaei, S. Abdolshah, Y. Akiyama, Y. Yamada and S. Okamoto are with Department of Mechanical Science and Engineering, Nagoya University, Nagoya, Japan (e-mail: nader.rajaei@mae.nagoya-u.ac.jp).

double-layer flooring can further reduce the impact force of. According to the best of the author's knowledge, the strategy of falling on a double-layer flooring has not been investigated.

This rest of this paper is organized as follows. In Section II, we introduce the method and protocol to be used for conducting the experiments. Section III represents the average results associated with impact force during the fall on a single layer flooring and the fall on double-layer flooring. The paper continues with further discussion and limitations regarding this study in section IV. Finally, the conclusions are presented in section V.

## II. MATERIAL AND METHOD

### A. Subjects

Four young Japanese individuals (male) of ages 20-25 years (mean ± standard deviation [SD] = 23.75 ± 0.5 years), participated in this study. Their average height and body mass were 174 ± 3 cm and 61 ± 5 kg, respectively. None of the subjects reported any prior training, such as Jodo, martial arts techniques, wrestling, or gymnastics. Moreover, they did not have a history of major medical or neurological illnesses, such as epilepsy, hand trauma, balance disorders, and falls. The participants did not present bone disease or other conditions that would increase the likelihood of a fracture occurrence. All subjects signed a written informed consent prior to the experiments. The experimental protocol was approved by the Institutional Review Board of Nagoya University, Japan. All experiments were conducted in accordance with the approved guidelines.

### B. Experimental Protocol

The objective of this experiment was to measure the forces applied to each hand during a forward fall while the ground is covered by a single-layer and double-layer flooring.

The experiment was designed according to the protocol utilized in most previous forward fall studies [18]-[19], [28]-[29]. This protocol was designed to be adopted in the *in vitro* study design [14]. In this study, each subject wore an upper body harness (Fig.1. (a)). The subjects were instructed to place their knees on a soft pillow fixed to the ground and flex their knees, allowing the subject to lean forward against the harness. In this position, the thighs and the horizontal line parallel to the ground created an angle of 30° (Fig.1).

A rope connected to the harness lifted the subjects to an appropriate height from the ground. The subjects were instructed to maintain at a position with their elbows fully extended (the outstretched hand position). In this position, the angle between the subjects' arms and the vertical line was established at 15°. Two force plates (M3D Force plate, Tec Gihan Corp.) were fixed to the ground beneath the subjects' hands. Given that the surface of the force plates is considered as a rigid surface with approximately infinite stiffness, in the first experiment, the surface of the force plates was covered by a single layer of foam pad (i.e. a compliant material) with 1 cm thickness (Fig.1.(b)). In the second experiments, an additional wood surface (rigid layer) with 1 cm thickness was inserted on top of the foam pad to become a double-layer flooring (Fig.1.(b)).

In order to prevent a wrist fracture during implementing the experiments, we considered the amount of impact force applied to each hand for two short fall heights ($h$ = 3 cm and 6 cm). The fall height was adjusted using the distance between the palms and the upper surface of the foam pad and the wood layer for the first and second experiment, respectively (Fig.1(c)). Every subject completed twelve trials (two experiments × two fall heights × three repetitions = twelve trials).

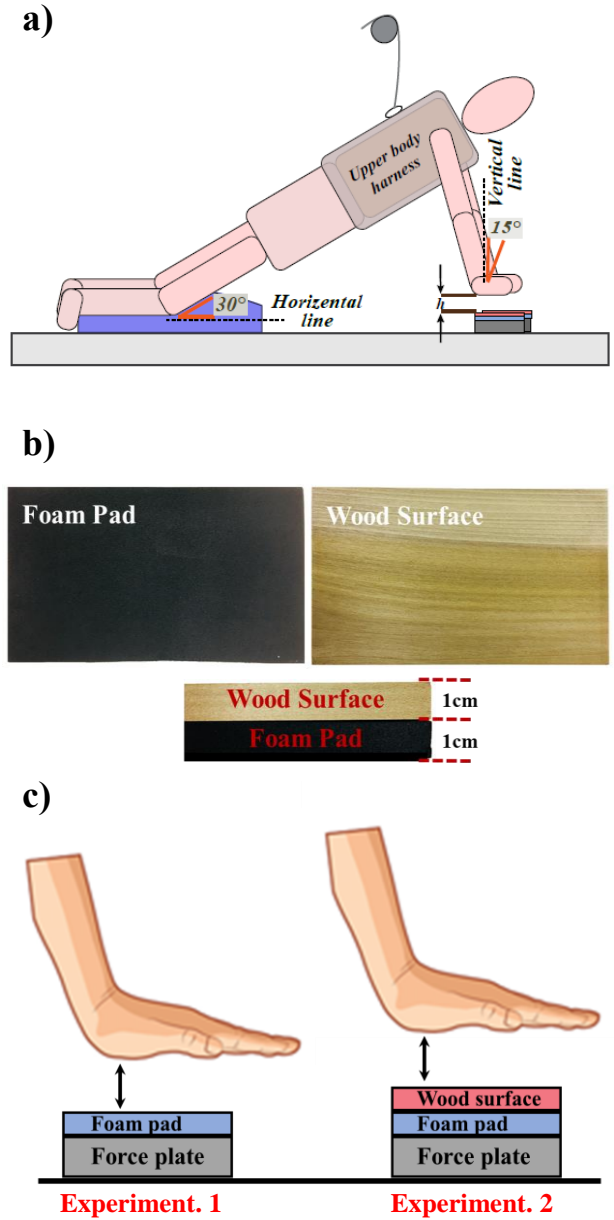

**Fig.1. (a)**. The experimental setup for the forward fall in the outstretched hand position for fall heights of 3 cm and 6 cm. **(b).** Two material layer, left side-foam pad, and right side-wood surface **(c)**. Different types of experiments: experiment 1 is associated when the force plates were covered by a single layer of the foam pad, experiment 2 is associated when one additional wood layer was fixed on the foam pad.

## III. RESULTS

Fig.2 illustrates the average impact force to be experienced by on hand for fall height of 3 cm (Fig.2. (a)) and 6 cm (Fig.2. (b)). Each plot consists of the result of both the experiment 1 (blue-dash line) and the experiment 2 (red-dash line). For both experiments, which is similar to the previous findings [18]-[21],[30] the impact force includes two peaks; the first one has a higher magnitude, which is followed by the lower second peak.

In Fig.2. (a), the average magnitude for the first peak was approximately measured as 278N and 250N for the experiment 1 and experiment 2, respectively. The magnitudes of the second peaks demonstrated lower forces of around 240N for experiment 1 and 243N for experiment 2. Once the fall height was increased to 6 cm (Fig.2. (b)), the average magnitude for the first and second peak in experiment 1 was approximately measured as 349N and 244N and in experiment 2 was 308N and 243N.

The one-way analysis of variance (ANOVA), using the SPSS software package (IBM Corp. Released 2016. IBM SPSS Statistics for Windows, Version 23.0. Armonk, NY: IBM Corp), illustrated that increasing the fall height from 3 cm to 6 cm significantly increases the magnitude of both peaks ($p < 0.00003$) for both experiment 1 and experiment 2. Additionally, increasing the fall height had a greater effect on the first peak than the second peak (Fig.3).

We also implemented other statistical analyses to compare the magnitude of impact force applied to the hand in experiment 1 and experiment 2 as shown in Fig.3 (two bar charts). The analyses indicated significantly attenuated in the magnitude of the first peak ($p < 0.01$ for height 3 cm (the yellow color-bar in Fig.3 (a)) and $p < 0.05$ for 6 cm (the yellow color-bar in Fig.3 (b))) in experiment 2 than experiment 1 whereas there was no significant effect on the magnitude of the second peak for both experiments ($p = 0.71$ for height 3 cm and $p = 0.95$ for height 6 cm (the blue color-bar in the Fig.3 (a) and (b)).

## IV. DISCUSSIONS

The previous biomechanical studies reported that using a compliant flooring is a potential method for attenuating the impact forces applied to the body during a fall [27]-[29],[31]. However, these studies have been focused on comparing the impact forces during fall on the ground with fall on one layer compliant surface [28]-[29]. In the present study, we investigated whether using a combination of a compliant layer and rigid layer (double-layer flooring) can reduce further the impact forces during a forward fall. In order to achieve the objective of this study, the impact forces for two different experiments were measured. Initially, the impact forces were measured when the ground was covered by a single layer of a compliant material (a foam pad). Next, the impact forces were again measured when one additional relatively rigid surface (a wood surface) was mounted on top of the first layer (double-layer surfaces). The experiments were conducted for two short fall heights (3cm and 6cm) to prevent the risk of fracture for the subjects during conducting experiments.

Our experimental results demonstrate that the profile of impact force for both experiments (experiment 1 and experiment 2) considerably are similar to the results conventionally claimed in previous studies [18]-[20],[30] and includes two peaks; the first peak consisted of a high magnitude which was followed by a second peak with a lower magnitude. The statistical analysis showed that the magnitude of the first peak was significantly reduced in experiment 2 than in experimental 1 (Fig.3). A calculation represented that the first peak in experiment 2 was reduced approximately 10.7% and 13.3% for fall height 3 cm and 6cm, respectively. Therefore, the magnitude of the first peak further decreased by increasing the fall height. Since the first peak impact force usually causes a fracture [32], the noticeable reduction of the first peak using a double-layer flooring including rigid layer and a compliant layer can be a vital point for preventing the wrist fractures within a forward fall from standing position.

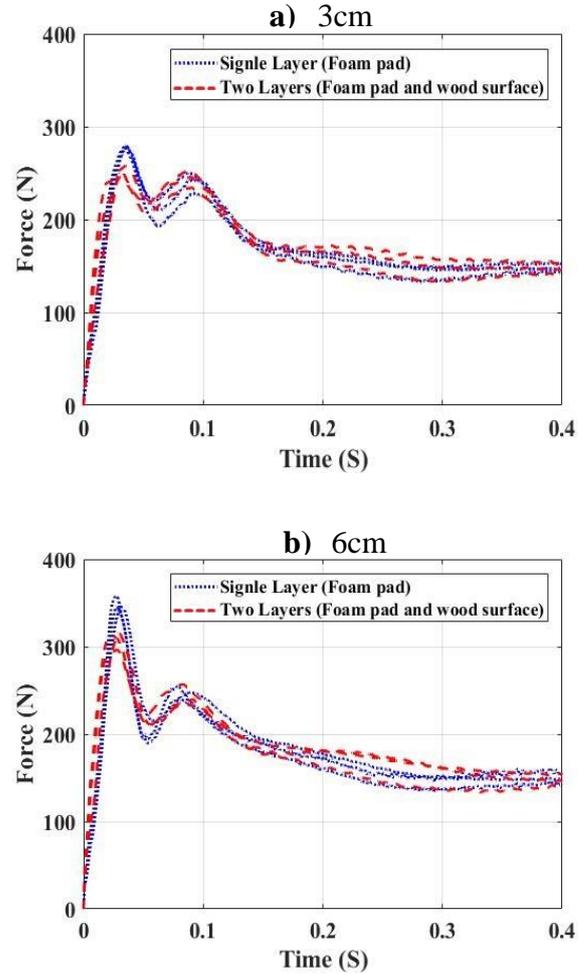

**Fig.2** The impact force profile during fall at **(a)** 3 cm and **(b)** 6cm. The figure consists of the average magnitude of force for experiment 1 (blue-dash line) and experiment 2 (red-dash line).

In addition, the experimental results indicated that increasing the fall height caused a significant increase in the magnitude of peaks. Therefore, the findings of this study provide additional evidence to support and confirm the previous findings regarding the impact force during a forward fall [18]-[20], [30].

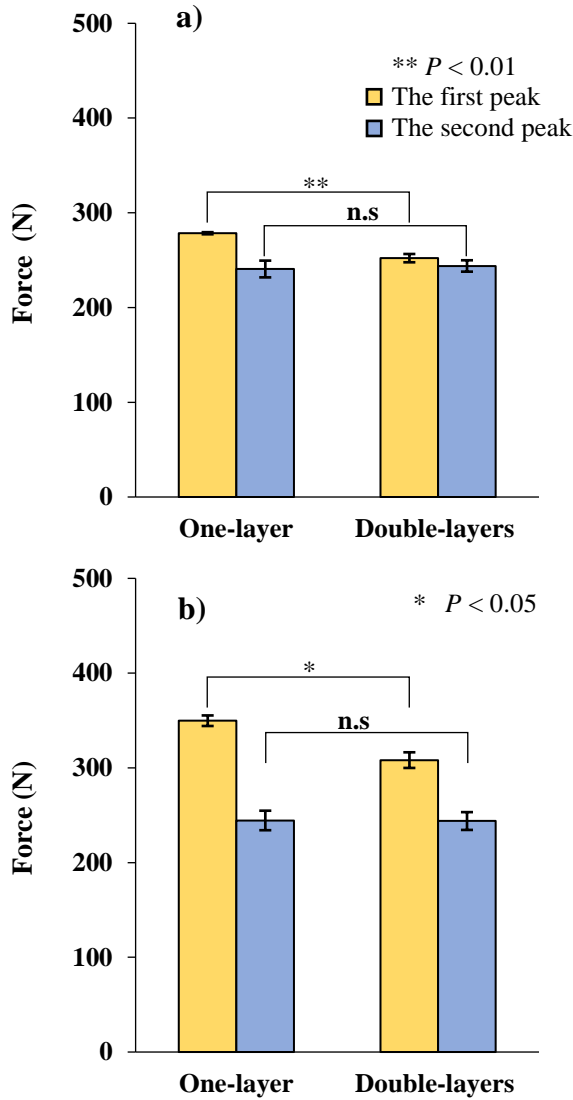

**Fig.3.** Comparison between the magnitude of the first peak ( yellow-bar) and the second peak (blue-bar) for one-layer and double-layer experiments during the fall from height (**a**) 3 cm and (**b**) 6 cm. *- significant for $p < 0.05$. **- significant for $p < 0.01$. **n.s-** no significant.

Furthermore, the experimental results of impact forces can be interpreted using a serially connected two spring model. According to Hook's law, the amount of compression force for a spring is calculated by multiplying the amount of reduction in the length of a spring to the amount of spring constant. Hence, each layer is modeled as a linear spring where the stiffness of each layer is considered as a spring constant and the amount of reduction in the length of springs is equivalent to the amount of compression of the layer during fall. In experiment 1, fall occurs on one layer (a foam pad), and the impact force is obtained from the behavior of a single spring (one degree of freedom). In contrast, in experiment 2, the impact force is calculated based on the behavior of two springs to be connected in series (two degrees of freedom) by considering double layer; one foam pad and one wood surface. In addition, the reduction of spring length is almost equal for both experiments because, in experiment 2, the reduction in the length of spring is incredibly small for the wood layer. Therefore, based on these assumptions, the amount of force in the experiment 2 must be smaller than experiment 1 because in experiment 2 the total amount of spring constant is calculated as if two-spring system connected in series whereas in experiment 1 there is a single spring constant. Subsequently, a simplistic linear spring system can well describe that fall on a double-layer flooring leads to a small amount of the impact forces applied to hand in comparison with a fall on the single compliant flooring.

However, some limitations are worth noting. Even though this study showed that the magnitude of impact force during a forward fall on the double-layer flooring decreases further in comparison with fall on a single compliant layer, more studies such as using a mathematical model or mathematical dynamic simulation are required to investigate results of using one-layer and double-layer flooring for a real forward fall from a typical standing position. Moreover, in this study, it was not investigated whether a double-layer flooring may increase the human balance during walking by decreasing the risk of fall as was reported in some previous studies [27], [33]-[34]. In our future work, this issue can be further investigated.

## V. CONCLUSIONS

In this study, for preventing the forward fall-related injuries, we demonstrated that using a double-layer flooring where a rigid layer is mounted on a compliant layer, could significantly reduce the impact force applied to the hand.

This result was obtained after conduction two laboratories experiments. In the first experiment, we measured the impact forces applied to the subjects' hands when they fell on one single compliant flooring from short fall heights. In the second experiment, we repeated the experiments where impact force was measured during the fall on the double-layer flooring.

In addition, our results illustrated similar impact force profiles for both experiments including two peaks; the first peak with a higher magnitude followed by a second peak with a lower magnitude of the force. More specifically, the double-layer flooring decreased the magnitude of the first peak which mainly causes a dorsal radius fracture during a forward fall. In contrast, the double-layer flooring not effected on the magnitude of the second peak.

The study results also showed that the magnitude of the first peak reduction is correlated with increase fall height so that by increasing the fall height from 3 cm to 6 cm, a further reduction was observed for the magnitude of the first peak. Hence, we can experimentally predict that fall form approximately large fall height, i.e. fall from a typical standing position, on a double-layer flooring remarkably decreases the first peak l. However, we have considered validating this prediction by appropriate mathematical models or mathematical dynamic simulation in our future studies.

## ACKNOWLEDGMENT

This work was supported by JSPS KAKENHI Grant Number 26750121, and METI "Strategic international standardization promotion project: International standardization of human tolerance against fall injuries".

REFERENCES

[1] Stenhagen, Magnus, Henrik Ekström, Eva Nordell, and Sölve Elmståhl. "Accidental falls, health-related quality of life and life satisfaction: a prospective study of the general elderly population." *Archives of gerontology and geriatrics* 58, no. 1, pp.95-100, 2014.

[2] Akyol, A. D. "Falls in the elderly: what can be done?." *International nursing review* 54, no. 2 (2007): 191-196.

[3] Gillespie, Lesley D., M. Clare Robertson, William J. Gillespie, Catherine Sherrington, Simon Gates, Lindy M. Clemson, and Sarah E. Lamb. "Interventions for preventing falls in older people living in the community." *Cochrane Database Syst Rev* 9, no. 11 (2012).

[4] von Heideken Wågert, Petra, Yngve Gustafson, Kristina Kallin, Jane Jensen, and Lillemor Lundin-Olsson. "Falls in very old people: the population-based Umeå 85+ study in Sweden." *Archives of gerontology and geriatrics* 49, no. 3 (2009): 390-396.

[5] Englander, Fred, Thomas J. Hodson, and Ralph A. Terregrossa. "Economic dimensions of slip and fall injuries." Journal of Forensic Science 41, no. 5, pp.733-746, 1996.

[6] Chang, Wen-Ruey, Sylvie Leclercq, Thurmon E. Lockhart, and Roger Haslam. "State of science: occupational slips, trips and falls on the same level." *Ergonomics* 59, no. 7, pp. 861-883, 2016.

[7] C. E. de Putter, E. F. van Beeck, S. Polinder, M. J. M. Panneman, A. Burdorf, S. E. R. Hovius, R. W. Selles, "Hand and Wrist Injuries by External Cause: A population-based study in working-age adults, 2008-2012," *Hand and Wrist Injuries*, vol. 47, no. 7, pp. 1478–1482, 2016.

[8] K. Mitsuoka, Y. Akiyama, Y. Yamada, and S. Okamoto, "Analysis of Skip Motion as a Recovery Strategy after an Induced Trip," in Systems, Man, and Cybernetics (SMC), 2015 IEEE International Conference on, pp. 911-916.

[9] T. W. O'Neill et al., "Age and sex influences on fall characteristics.," Ann. Rheum. Dis., vol. 53, no. 11, pp. 773–5, 1994.

[10] Nevitt, Michael C., and Steven R. Cummings. "Type of fall and risk of hip and wrist fractures: the study of osteoporotic fractures." Journal of the American Geriatrics Society 41, no. 11 (1993): 1226-1234.

[11] J. A. Baron et al., "Basic epidemiology of fractures of the upper and lower limb among Americans over 65 years of age.," *Epidemiology*, vol. 7, no. 6, pp. 612–8, 1996.

[12] L. M. Feehan and S. B. Sheps, "Incidence and Demographics of Hand Fractures in British Columbia, Canada: A Population-Based Study," J. Hand Surg. Am., vol. 31, no. 7, pp. 1–9, 2006.

[13] Nevitt, Michael C., and Steven R. Cummings. "Type of fall and risk of hip and wrist fractures: the study of osteoporotic fractures." Journal of the American Geriatrics Society 41, no. 11 (1993): 1226-1234.

[14] G. Frykman, "Fracture of the distal radius including sequelae--shoulder-hand-finger syndrome, disturbance in the distal radio-ulnar joint and impairment of nerve function. A clinical and experimental study.," Acta Orthop. Scand., vol. 6470, no. April, p. Suppl 108:38, 1967

[15] E. R. Myers et al., "Correlations between photon absorption properties and failure load of the distal radius in vitro," Calcif Tissue Int, vol. 49, no. October 2016, pp. 292–297, 1991.

[16] E. R. Myers, A. T. Hecker, D. S. Rooks, J. A. Hipp, and W. C. Hayes, "Geometric variables from DXA of the radius predict forearm fracture load in vitro," Calcif. Tissue Int., vol. 52, no. 3, pp. 199–204, 1993.

[17] J. A. Spadaro, F. W. Werner, R. A. Brenner, M. D. Fortino, L. A. Fay, and W. T. Edwards, "Cortical and trabecular bone contribute strength to the osteopenic distal radius," J. Orthop. Res., vol. 12, no. 2, pp. 211–218, 1994.

[18] J. Chiu and S. N. Robinovitch, "Prediction of upper extremity impact forces during falls on the outstretched hand," Journal of Biomechanics, vol. 31, no. 12, pp. 1169–1176, 1998.

[19] C. E. Kawalilak, J. L. Lanovaz, J. D. Johnston and S. A. Kontulainen, "Linearity and sex-specificity of impact force prediction during a fall onto the outstretched hand using a single-damper-model," J Musculoskelet Neuronal Interact, vol. 14, no. 3, pp. 286-293, 2014.

[20] DeGoede, Kurt M., and James A. Ashton-Miller. "Biomechanical simulations of forward fall arrests: effects of upper extremity arrest strategy, gender and aging-related declines in muscle strength." Journal of biomechanics 36, no. 3, pp. 413-420, 2003.

[21] Lo, JiaHsuan, and James A. Ashton-Miller. "Effect of upper and lower extremity control strategies on predicted injury risk during simulated forward falls: a study in healthy young adults." Journal of biomechanical engineering 130, no. 4, pp. 041015, 2008a.

[22] S. Abdolshah, Y. Akiyama, K. Mitsuoka, Y. Yamada and S. Okamoto, "Analysis of Upper Extremity Motion during Trip-Induced Falls," in Robot and Human Interactive Communication (RO-MAN), 26th IEEE International Symposium on Robot and Human Interactive Communication, pp. 1485-1490, 2017.

[23] K. M. DeGoede and J. A. Ashton-Miller, "Fall arrest strategy affects peak hand impact force in a forward fall," J. Biomech., vol. 35, no. 6, pp. 843–848, 2002.

[24] P. H. Chou, Y. L. Chou, C. J. Lin, F. C. Su, S. Z. Lou, C. F. Lin, and G. F. Huang, "Effect of elbow flexion on upper extremity impact forces during a fall," Clinical Biomechanics, vol. 16, pp. 888–894, 2001.

[25] P. P-H Chou, H-C Chen, H-H Hsu, Y-P Huang, T-C Wu and Y-L Chou, "Effect of upper extremity impact strategy on energy distribution between elbow joint and shoulder joint in forward falls," Journal of Medical and Biological Engineering, vol. 32, no. 3, pp. 175-180, 2012.

[26] J. Lo, G. N. McCabe, K. M. DeGoede, H. Okuizumi, and J. A. Ashton-Miller, "On reducing hand impact force in forward falls: results of a brief intervention in young males," Clinical biomechanics, vol. 18, no. 8, pp. 730-736, 2003.

[27] Lachance, Chantelle C., Michal P. Jurkowski, Ania C. Dymarz, Stephen N. Robinovitch, Fabio Feldman, Andrew C. Laing, and Dawn C. Mackey. "Compliant flooring to prevent fall-related injuries in older adults: A scoping review of biomechanical efficacy, clinical effectiveness, cost-effectiveness, and workplace safety." PLoS one 12, no. 2 , pp. e0171652, 2017.

[28] S. N. Robinovitch and J. Chiu, "Surface stiffness affects impact force during a fall on the outstretched hand," Journal of Orthopaedic Research, vol. 16, no. 3, pp. 309–313, 1998.

[29] Choi, Woochol Joseph, Harjinder Kaur, and Stephen N. Robinovitch. "Effect of Playground Surface Materials on Peak Forces during Falls." CMBES Proceedings 34, no. 1, 2018.

[30] J. A. Ashton-Miller, "Biomechanical Factors Affecting the Peak Hand Reaction Force During the Bimanual Arrest of a Moving Mass," J. Biomech. Eng., vol. 124, no. 1, p. 107, 2001.

[31] Lachance, Chantelle C., Valeriya O. Zaborska, Pet-Ming Leung, Fabio Feldman, Stephen N. Robinovitch, and Dawn C. Mackey. "Perceptions about Compliant Flooring from Senior Managers in Long-Term Care." Journal of Housing For the Elderly, pp.1-17, 2018.

[32] Gustavsson J. Working in a nursing home with Impact Absorbing Flooring: A qualitative study on the experiences of licensed practical nurses [Internet]. Doctoral Dissertation, Karlstad University. 2015.

[33] Simpson, A. H. R. W., S. Lamb, P. J. Roberts, T. N. Gardner, and J. Grimley Evans. "Does the type of flooring affect the risk of hip fracture?.." Age and Ageing 33, no. 3, pp.242-246,2004.

[34] Warren, Christopher J., and Hugh C. Hanger. "Fall and fracture rates following a change from carpet to vinyl floor coverings in a geriatric rehabilitation hospital. A longitudinal, observational study." Clinical rehabilitation 27, no. 3, pp.258-263, 2013.